\documentclass[12pt]{iopart}

\usepackage{graphicx}
\usepackage{dcolumn}
\usepackage{amsgen}
\usepackage{amsfonts}
\usepackage{amssymb}
\usepackage{amsbsy}
\usepackage{bm}
\usepackage{color}

\usepackage{epstopdf}

\begin{document}

\newcommand{\beq}{\begin{equation}}
\newcommand{\eeq}{\end{equation}}
\newcommand{\barr}{\begin{eqnarray}}
\newcommand{\earr}{\end{eqnarray}}

\newcommand{\andy}[1]{ }

\newcommand{\bmsub}[1]{\mbox{\boldmath\scriptsize $#1$}}

\def\bra#1{\langle #1 |}
\def\ket#1{| #1 \rangle}
\def\sinc{\mathop{\text{sinc}}\nolimits}
\def\cV{\mathcal{V}}
\def\cH{\mathcal{H}}
\def\cT{\mathcal{T}}
\def\cP{\mathcal{P}}
\renewcommand{\Re}{\mathop{\text{Re}}\nolimits}
\renewcommand{\Im}{\mathop{\text{Im}}\nolimits}

\newcommand{\rev}[1]{{\color{red}#1}}
\newcommand{\REV}[1]{\textbf{\color{red}#1}}
\newcommand{\BLUE}[1]{\textbf{\color{blue}#1}}
\newcommand{\GREEN}[1]{\textbf{\color{green}#1}}

\title{Statistical mechanics of multipartite entanglement}

\author{P. Facchi$^{1,2}$, G. Florio$^{3,2}$, U. Marzolino$^{4}$, G. Parisi$^{5}$, S. Pascazio$^{3,2}$}

\address{$^1$Dipartimento di Matematica, Universit\`a di Bari, I-70125 Bari, Italy}
\address{$^2$Istituto Nazionale di Fisica Nucleare, Sezione di Bari, I-70126 Bari, Italy}
\address{$^3$Dipartimento di Fisica, Universit\`a di Bari, I-70126 Bari, Italy}
\address{$^4$Dipartimento di Fisica Teorica, Universit\`a di Trieste, Strada Costiera
11, 34014 Trieste, Italy and INFN, Sezione di Trieste, Trieste,
Italy}
\address{$^5$Dipartimento di Fisica, Universit\`{a} di Roma ``La Sapienza", Piazzale Aldo Moro 2, 00185 Roma,
Italy, Centre for Statistical Mechanics and Complexity (SMC),
CNR-INFM, 00185 Roma, Italy and INFN, Sezione di Roma, 00185 Roma,
Italy}

\date{\today}

\begin{abstract}
We characterize the multipartite entanglement of a  system of $n$ qubits
in terms of the distribution function of the bipartite
purity over all balanced bipartitions. We search for those (maximally
multipartite entangled) states whose purity is minimum for
all bipartitions and recast this optimization problem into a problem
of statistical mechanics.
\end{abstract}


\maketitle

Entanglement is a consequence of the quantal superposition principle
and therefore of the linearity of Hilbert spaces, and embodies the
impossibility of factorizing a given state of a total quantum system
in terms of the states of its constituents. The notion of
entanglement plays a central role in the development of novel
resources and techniques in computing, communication and information
processing. There is a widespread literature about the
characterization, quantification and properties of bipartite
entanglement, i.e.\ the entanglement of \emph{two} subsystems
\cite{entanglement,entanglement1}. The landscape of \emph{multipartite}
entanglement is less understood but very widely investigated
\cite{multipart,multipart1,multipart2,multipart3}. The main difficulty is that there
is no unique way of quantifying it: more so, different definitions
often do not agree with each other, because they adopt different
strategies and capture different features of this inherently quantum
phenomenon.

It would be desirable to understand how to properly characterize
multipartite entanglement, e.g.\ by identifying a few key properties
that can account for its overall features. The quantification and
evaluation of global entanglement measures bears serious
computational difficulties, because states endowed with large
entanglement typically involve exponentially many coefficients.
Multipartite entanglement has therefore the typical features of a
complex phenomenon, and for this reason we shall follow a precept
derived from the study of complex systems, that relies on the idea
that complicated phenomena cannot be summarized by a single (or a
few) number(s) \cite{parisi}. We notice that in the context of
quantum entanglement this idea was alluded to in \cite{MMSZ}. In
this spirit, entanglement can be analyzed in terms of a function:
the probability density of the purity of a subsystem over all
bipartitions of the total system
\cite{FFP,FFP1}. A state has a large \emph{multipartite}
entanglement if it has a large and well distributed \emph{bipartite}
entanglement. In other words, states with large multipartite
entanglement are characterized by a narrow distribution of bipartite
entanglement, centered around a large average.

In this letter we shall explore these ideas even further and
endeavor to give a statistical mechanical characterization of
multipartite entanglement, by recasting the problem in terms of a
\emph{cost function} that will represent the (inverse) global
entanglement of a quantum system. We will establish a direct
connection between the study of multipartite entanglement and the
tools of classical statistical mechanics.

We shall consider an ensemble of $n$ spin-1/2 particles (qubits),
whose Hilbert space $\mathcal{H}=(\mathbb{C}^2)^{\otimes n}$ has
dimension $N=2^n$. We divide this set in two parts, $A$ and
$\bar{A}$, made up of $n_A$ and $n_{\bar{A}}$ qubits respectively
$(n_A + n_{\bar{A}}=n)$. The total Hilbert space is split in the
tensor product
$\mathcal{H}=\mathcal{H}_A\otimes\mathcal{H}_{\bar{A}}$, with
dimensions $N_A=2^{n_A}$ and $N_{\bar{A}}=2^{n_{\bar{A}}}$
respectively $(N_AN_{\bar{A}}=N)$. We also assume with no loss of
generality $n_A\leqslant n_{\bar{A}}$. We shall consider only pure
states and shall not discuss additional phenomena such as bound
entanglement \cite{boundent,boundent1}.
Let
\begin{equation}
|\psi\rangle = \sum_{j=0}^{N-1} z_j |j\rangle , \quad z_j \in
\mathbb{C}, \quad \sum_{j=0}^{N-1} {|z_j|}^2 =1,
\label{eq:genrandomx}
\end{equation}
where the computational base $\left\{|j\rangle\right\}$ is
expressed in terms of the  eigenstates of the third Pauli matrices
acting on each site. These yield binary sequences belonging to
$\left\{0,1\right\}^{n}$, whose decimal representation are the
indices $j$. For each bipartition $(A,\bar{A})$ we define a
bijection $j\leftrightarrow(j_A,j_{\bar{A}})$, where the integers
$j_A\in[0,N_A-1]$ and $j_{\bar{A}}\in[0,N_{\bar{A}}-1]$ are the
decimal representation of the subsequences of portion $A$ and
${\bar{A}}$, respectively. This reflects the factorization
$|j\rangle=|j_A\rangle\otimes|j_{\bar{A}}\rangle$. As bipartite
entanglement measure [at fixed bipartition $(A,{\bar{A}})$] we take
the purity of subsystem $A$, which is conveniently expressed in
terms of the reduced density operators, $\rho_A=\tr_{\bar{A}}\rho$
($\rho=|\psi\rangle\langle\psi|$ being the total density operator)
\begin{eqnarray}
 \pi_{A}&=&\tr_A\rho_A^2=\tr_{\bar{A}}\rho_{\bar{A}}^2
=\sum_{j_A,l_A,j_{\bar{A}},l_{\bar{A}}}z_{j_A,j_{\bar{A}}}\bar{z}_{l_A,j_{\bar{A}}}
z_{l_A,l_{\bar{A}}}\bar{z}_{j_A,l_{\bar{A}}} .
\label{purity}
\end{eqnarray}
Purity ranges in $\left[1/N_A,1\right]$ and is $1$ if and only if
$\rho_A$ and $\rho_{\bar{A}}$ are projectors, that is, if the state
is (bi)separable,
$|\psi\rangle=|\psi_A\rangle\otimes|\psi_{\bar{A}}\rangle$.
Otherwise $|\psi\rangle$ is entangled. $\pi_{A}$ saturates its
minimum $1/N_A$ if and only if the reduced density operator of the
smaller partition is proportional to the identity matrix
$\rho_A=\mathbf{1}/N_A$.
The latter is the case of states $|\psi\rangle$ endowed with maximum
bipartite entanglement, at \emph{fixed} bipartition $(A,\bar{A})$.
The distribution function of purity over all
bipartitions, $p(\pi_{A})$, yields a probability-density-function
characterization of multipartite entanglement: its average measures
the mean entanglement of the state when the bipartitions are varied,
while its variance quantifies how uniformly bipartite entanglement
is distributed among all possible bipartitions.

The coefficients $z_j$ in (\ref{eq:genrandomx}) belong to the set
\begin{equation}\label{eq:constraint}
C=\left\{(z_0,z_1,\ldots,z_{N-1})\in\mathbb{C}^N | \sum_k
|z_k|^2=1\right\}.
\end{equation}
Note that $C$ corresponds to the set of normalized vectors
$\mathcal{S}=\{\psi\in\mathcal{H}\, |\, \|\psi\|=1\}$ and is  left
invariant under the natural action of the unitary group
$\mathcal{U}(\mathcal{H})$. A typical state is obtained by a uniform
sampling of $\mathcal{S}$. Typical states can be (efficiently)
generated by a chaotic dynamics \cite{chaos,chaos1}.

In the limit of large $N$, the $\pi_A$'s have a bell shaped
distribution over the bipartitions with mean and variance
\barr
\mu &=&\langle \pi_{A}\rangle
= \frac{N_A+N_{\bar{A}}}{N+1}, \label{eq:6a} \\
\sigma^2 &=&
\langle (\pi_{A}-\mu)^2\rangle=\frac{2(N_A^2-1)(N_{\bar{A}}^2-1)}{(N+1)^2(N+2)(N+3)} , \label{eq:6b}
\earr
respectively \cite{FFP,FFP1,aaa,aaa1,aaa2,aaa3,aaa4,aaa5}. The
brackets $\langle
\cdots
\rangle$
denote the average with respect to the unitarily invariant measure
over pure states
\begin{equation}
d\mu_C(z)=\prod_k dz_k d\bar{z}_k \delta \! \left(1-\sum_k
|z_k|^2\right),
\label{eq:meastyp}
\end{equation}
induced by the Haar measure over $\mathcal{U}(\mathcal{H})$ through
the mapping $|\psi\rangle = \sum  z_j |j\rangle = U |\psi_0\rangle$,
for a given state $|\psi_0\rangle$ \cite{aaa3}.

\begin{figure}
\centering
\includegraphics[width=0.8\textwidth]{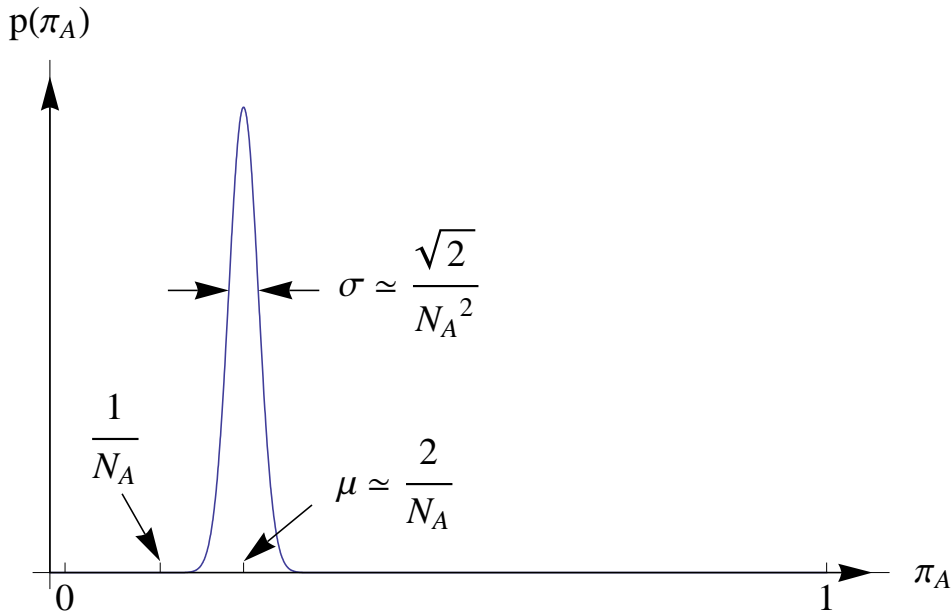}
\caption{Probability density function
$p(\pi_{A})$ of a typical state over balanced bipartitions. For
large $n$, the average is $2/N_A$ and the standard deviation
$\sqrt{2}/N_A^2$ ($N_A\simeq \sqrt{N}$). Our objective is to
characterize those ``maximally multipartite entangled states"
whose average purity is minimum.}
\label{probpurity6arrow}
\end{figure}

It is interesting to notice that, for balanced bipartitions
$n_A=\lfloor n/2 \rfloor$, the average and standard deviation of the
distribution simplify to $\mu \simeq 2 /N_A, \sigma \simeq \sqrt{2}
/N_A^2$, the approximation becoming quickly very accurate as the
number of qubits increases. The behavior of the probability density
function $p(\pi_{A})$ is sketched in Fig.\
\ref{probpurity6arrow}. It is worth emphasizing that the typical
states are very entangled, but are far from the minimum (on average)
by a factor 2; moreover, their distribution function is very narrow
(a signature of a good multipartition of entanglement), but has a
nonvanishing standard deviation. It is then natural to ask: Is it
possible to find states that perform better? The answer to this
question is positive \cite{MMES}: there exist
``maximally multipartite entangled states" (MMES) such that their
distribution function (over balanced bipartitions) is a Dirac delta
function. For $n=2,3,5,6$ the delta is centered on the
\emph{smallest} possible value: $p(\pi_{A})=\delta(\pi_{A}-1/N_A)$.
For $n=2$ MMES are Bell states up to local unitary
transformations, while for $n=3$ MMES are equivalent to the GHZ states
\cite{GHZ}.
For $n=4$ one numerically obtains $p(\pi_{A})=\delta(\pi_{A}-C_4)$,
with $C_4 = 1/3 > 1/4 =1 /N_A$ (but still $C_4 < 1/2=2 /N_A$)
\cite{MMES,sudbery,sudbery1,sudbery2}. For $n=7$ one finds states such that
$p(\pi_{A})=\delta(\pi_{A}-C_7)$, with $C_7 \simeq 0.136
\gtrsim 1/8 = 1 /N_A$.
``Perfect" MMES (that saturate the minimum purity $1/N_A$)
do not exist for $n\geq 8$ \cite{scott}. This will be presently
viewed as a symptom of
\emph{frustration} and is a new and intriguing feature of the
multipartite scenario, as opposed to the bipartite case. We shall
now endeavor to outline a strategy that enables one to
characterize MMES.

The search for MMES can be recast in terms of an optimization
problem. Look at the average purity as a cost function
(``potential of multipartite entanglement"\cite{MMES})
\begin{equation} \label{hamiltonian}
H=\mathbb{E}[\pi_A] \equiv \left(\begin{array}{l}n
\\n_A\end{array}\!\!\right)^{-1}\sum_{|A|=n_A}\pi_{A},
\end{equation}
where $\mathbb{E}$ denotes the expectation value and the sum is over
balanced bipartitions $n_A=\lfloor n/2
\rfloor$. Since we are focusing on
balanced bipartitions, and any bipartition can be brought into any
other bipartition by applying a permutation of the qubits, the sum
over balanced bipartitions in (\ref{hamiltonian}) is equivalent to a
sum over the permutations of the qubits. $H$ in Eq.\ (\ref{hamiltonian}) is related to the linear entropy considered in \cite{scott,linear.entropy}.

The above minimization problem has proved to be of formidable
difficulty, mainly for the presence of frustration: the
requirement that purity saturates its minimum $1/N_A$ can be
satisfied for some but \emph{not} for \emph{all} balanced
bipartitions. In other words, not all terms in the summation
(\ref{hamiltonian}) can be $1/N_A$. In order to study this
problem, we will recast it in \emph{classical statistical
mechanical} terms: instead of searching for the minimum of $H$,
one can look at the free energy of a suitable classical system at a
fictitious temperature and recover the original problem in the
zero temperature limit.

We follow the standard ensemble approach of statistical
mechanics and introduce an ensemble $\{m_{j}\}$ of $M$ vectors
(states), where $m_j$ is the number of vectors with purity
$H=\epsilon_j$. One seeks the distribution that maximizes the number
of states $\Omega=M!/\prod_j m_{j}!$ under the constraints that
$\sum_j m_{j} =M$ and $\sum_j m_{j}\epsilon_j =ME$. For $M \to
\infty$, the above optimization problem yields the canonical ensemble
and its partition function
\begin{equation} \label{partition.function}
Z(\beta) = \int d\mu_C(z) e^{-\beta H(z)},
\end{equation}
where the Lagrange multiplier $\beta$, that plays the role of an
inverse temperature, fixes the average value of purity $E$. The
constraint $C$ in the integration domain is due to the normalization
(\ref{eq:constraint}) and the measure is defined in
(\ref{eq:meastyp}). Note that the partition function
(\ref{partition.function}) can be given a base-independent
expression
\begin{equation}
\fl\qquad Z(\beta) = \int d\mu_C(z) e^{-\beta H(z)}= c_N \int d\mu_{\mathrm{H}}(U)
\exp\left(-\beta \mathbb{E}[\tr_A (\tr_{\bar{A}} U\ket{\psi_0}\bra{\psi_0}U^\dagger)^2]\right),
\end{equation}
where $\mu_{\mathrm{H}}$ denotes the Haar measure over
$\mathcal{U}(\mathcal{H})$, $\ket{\psi_0}$ is any given vector and
the (unimportant) constant
$c_N=\mu_C(\mathbb{C}^N)/\mu_{\mathrm{H}}(\mathcal{U}(\mathcal{H}))$
is proportional to the ratio between the area of the
$(N-1)$-dimensional sphere (\ref{eq:constraint}) and the volume of
the unitary group.

The average over bipartitions in (\ref{hamiltonian}) entails a
complicated dependence on the indices $j$ in Eqs.
(\ref{eq:genrandomx})-(\ref{purity}). This hinders one from finding
a closed expression for the partition function and makes the
application of a standard saddle-point method or the calculation of
the quenched average rather involved. Nonetheless, by applying the
standard tools of statistical mechanics, one can discuss several
aspects and analyze interesting limits. For $\beta\to0$ Eq.\
(\ref{partition.function}) yields the distribution of the typical
states (\ref{eq:meastyp}). Notice that this is valid both for $\beta
\to 0^\pm$. For $\beta\to +\infty$, only those configurations that
minimize the Hamiltonian survive, namely the MMES. Remarkably, there
is a physically appealing interpretation even for negative
temperatures: for $\beta\to -\infty$, those configurations are
selected that maximize the Hamiltonian, that is separable states.
These preliminary findings are summarized in Table
\ref{betalims}.
\begin{table}
\caption{High, low and negative temperature limits. }
\label{betalims}
\begin{tabular}{|l|l|l|}
\hline
  $\beta\to +\infty$ & $H=E_0$ (min) & MMES  \\
  $\beta\to 0 $ & $H\simeq\mu$ & typical states  \\
  $\beta\to -\infty $ & $H=1$ (max) & separable states \\
 \hline
\end{tabular}
\end{table}

\begin{figure}
\centering
\includegraphics[width=0.8\textwidth]{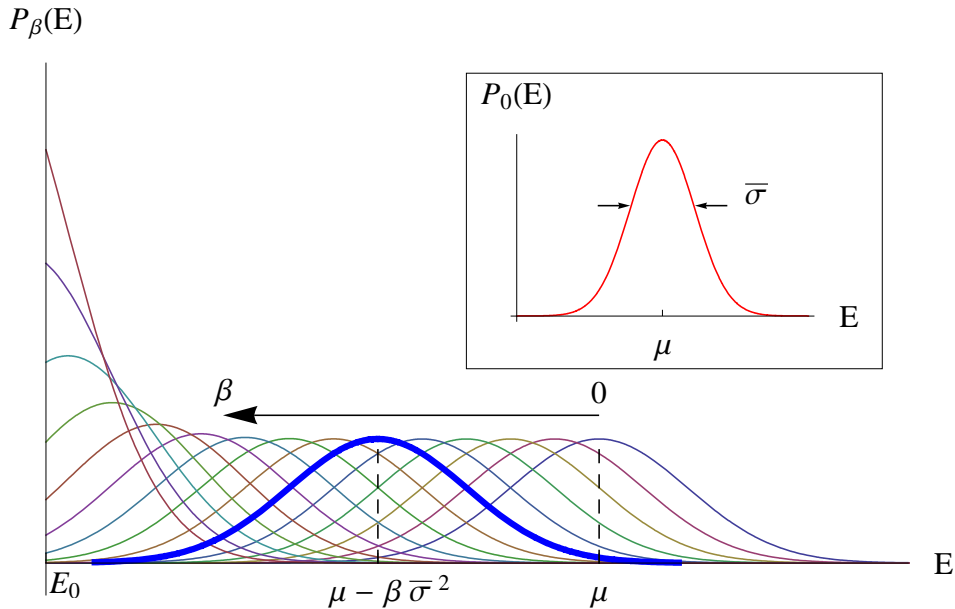}
\caption{
Qualitative sketch of Eq.\ (\ref{energy.distribution}), at fixed
$N_A$,  in arbitrary units. The energy  density
function is distributed around $\mu$ with standard deviation $\bar{\sigma}$ at $\beta=0$ (inset)
and moves toward $E_0$ when $\beta\to +\infty$.
From right to left, $\beta$ changes in constant steps.
The probability density rigidly shifts with $\beta$, for $\beta
\lesssim N^{7/2-\log_2 3}$. See Eq.\ (\ref{eq:shift}) and
following discussion. Note that $E_0=O(N^{-1/2})$. }
\label{figtdependence}
\end{figure}

By manipulating the partition function we can write the energy
distribution function at arbitrary $\beta$ in terms of its high
temperature limit:
\begin{eqnarray} \label{energy.distribution}
P_\beta(E)&=&\frac{1}{Z(\beta)} \int d\mu_C(z)
\delta\left(H-E\right) e^{-\beta H}
=\frac{e^{-\beta E}P_0(E)}{\int_{E_0}^1 dE e^{-\beta E}P_0(E)},
\end{eqnarray}
where $E\in[E_0,1]$, $E_0$ being the minimum of the potential of multipartite entanglement
$H$ in (\ref{hamiltonian}). We know that  $1/N_A\leq E_0 (N_A) \leq \mu\leq 2/N_A$,
and thus $\lim_{N_A \to \infty} E_0
(N_A)=0$ . By multiplying and dividing the
last equation by $|\beta|e^{\beta}$ and $\beta e^{\beta E_0}$,
respectively, we find
\begin{equation} \label{betalimits}
P_{-\infty}(E)=\delta(E-1), \qquad P_{+\infty}(E)=\delta(E-E_0).
\end{equation}
These limits are the counterparts of those discussed for the
partition function and are reflected in the asymptotic behaviour of
the average energy as function of $\beta$
\begin{eqnarray} \label{mean.energy}
\langle H\rangle_\beta&=&\frac{1}{Z(\beta)} \int d\mu_C(z)
H e^{-\beta H} =\int_{E_0}^1 dE E P_\beta(E)=-\frac{\partial}{\partial\beta}\ln
Z(\beta).
\end{eqnarray}
Indeed,
\begin{equation} \label{Hbetalimits}
\langle
H\rangle_{-\infty}=1, \quad \langle H\rangle_{+\infty}=E_0 .
\end{equation}
Moreover $\frac{\partial}{\partial\beta}\langle
H\rangle_\beta=-\langle H^{2}\rangle_\beta+\langle
H\rangle_\beta^{2} \equiv -\triangle H^2_\beta \leqslant 0$, which
is nonpositive. Thus the average energy is a non-increasing
function of $\beta$ and has at least one inflexion point as function
of $\beta$. From a qualitative point of view, one expects the
behavior sketched in Fig.\ \ref{figtdependence}: for $\beta
\to 0 $, the distribution is Gaussian
(typical states); when $\beta\to +\infty$ the
distribution tends to become more concentrated around $E_0$. This
picture will be substantiated in the following.

In general, the high-temperature expansion of the average energy
reads
\begin{eqnarray}
\label{high.temp}
\langle H\rangle_\beta &=&
\sum_{m=1}^\infty\frac{(-\beta)^{m-1}}{(m-1)!}\kappa_{0}^{(m)}(H) ,
\\
\label{derivatives}
\kappa^{(m)}_\beta(H) &=&
(-)^{m-1}
\frac{\partial^{m-1}}{\partial\beta^{m-1}}\langle H\rangle_\beta ,
\end{eqnarray}
where $\kappa_{\beta}^{(m)}(H)$ are the cumulants of $H$. So far,
the analysis is valid at fixed $N=2^n$. Let us now briefly discuss
the limit $N \to \infty$. The first cumulant is
nothing but the average purity $\mu$ defined in (\ref{eq:6a}). The
second cumulant
 $\bar{\sigma}^2= \triangle H^2_0=\kappa_{0}^{(2)}(H)$ can
 also be computed exactly. A detailed calculation, to be presented
 elsewhere, yields
\begin{equation} \label{eq:secondocum}
\bar{\sigma}^2= \kappa_{0}^{(2)}(H) \sim 3\sqrt{2}N^{-4+\log_2
3} \simeq O (N^{-2.42}).
\end{equation}
The presence of an irrational exponent
is a rather rare feature and an interesting phenomenon in itself.
We notice that if the $\pi_A$'s in the sum (\ref{hamiltonian})
were independent Gaussian random variables, one would get
$\bar{\sigma}^2\sim \sigma^2/N =O(N^{-3})$, where $\sigma$ is
defined in Eq.\ (\ref{eq:6b}) (the number of balanced bipartitions
being of order $N$.) This shows that different bipartitions
``interfere," yielding a highly nontrivial average. In fact, it
can be shown that the correlations decrease exponentially with the
distance between bipartitions.

Higher order cumulants at $\beta=0$ can be shown to tend to zero
faster than the others. Since the Gaussian is the only probability
distribution with a finite number of non-vanishing cumulants (the
first and the second one), one can therefore assume a Gaussian
approximation for the energy distribution at infinite temperature,
by neglecting all the cumulants but the first two:
\begin{equation} \label{gauss.approx}
P_{0}(E)\sim \frac{1}{\sqrt{2\pi\bar{\sigma}^2}}\exp\left(-\frac{\left(E-\mu\right)^{2}}{2\bar{\sigma}^2}\right),
\end{equation}
where $\mu$ and $\bar{\sigma}$ are defined in (\ref{eq:6a}) and
(\ref{eq:secondocum}), respectively. Finally, by applying
(\ref{energy.distribution}) we can compute the energy distribution
at arbitrary temperature
\begin{eqnarray} \label{gauss.approx2}
P_{\beta}(E) & \sim & \frac{1}{\sqrt{2\pi\bar{\sigma}^2 }}
\exp\left(-\frac{\left(E-\mu+\beta \bar{\sigma}^2 \right)^{2}}
{2 \bar{\sigma}^2}\right) 
\end{eqnarray}
Thus from the definition (\ref{derivatives}) of cumulants, $\langle
H\rangle_\beta$ as a function of $\beta$ would have a vanishing
curvature, in contradiction with the qualitative sketch mentioned
after Eq.\ (\ref{mean.energy}).
Eq.\ (\ref{gauss.approx2}) can be written as
\begin{eqnarray} \label{eq:shift}
P_{\beta}(E) = P_0(E+\beta \bar{\sigma}^2),
\end{eqnarray}
an expression valid also for nongaussian distributions that are well
concentrated around their mean $\mu$. As a matter of fact, Eqs.\
(\ref{gauss.approx2}) and (\ref{eq:shift}) are valid only when
$\beta$ is not too large. In fact, they hold true as far as
$\mu-\beta\bar{\sigma}^2-\bar{\sigma}\gtrsim0$, i.e.\
$\beta\lesssim\mu/\bar{\sigma}^2 =O( N^{7/2-\log_2 3})\simeq
O(N^{1.92})$. Up to this value the probability density rigidly
shifts with $\beta$, as is apparent in Fig.\ \ref{figtdependence}.
For larger values of $\beta$ the lower tail of the distribution
starts ``feeling" the wall at $E_0$. The large-$\beta$ asymptotic form of
$P_\beta(E)$ depends on the behavior of $P_0(E\rightarrow E_0)$: one easily obtains
\begin{equation}\label{eq:betalarge}
P_\beta(E)\sim \frac{\beta^{\ell+1}}{\ell!}(E-E_0)^\ell e^{-\beta(E-E_0)},
\end{equation}
where $\ell$ is the order of the first nonvanishing derivative of
$P_0(E)$ at $E_0$. (Figure \ref{figtdependence} displays the case
$\ell=0$.) Notice that the only relic of $P_0(E)$ in
(\ref{eq:betalarge}) is $\ell$ and $P_{\beta \to +\infty}(E)$ yields the
second equation in (\ref{betalimits}). The analysis for
$\beta\to-\infty$ is analogous and yields the first equation in
(\ref{betalimits}).

In conclusion, we have built a statistical theory around an
optimization problem, whose solutions are the maximally
multipartite entangled states, that appear as minimal energy
configurations.
The approach we propose borrows methods from
classical statistical mechanics in order to investigate the
multipartite entanglement scenario. The introduction of a temperature
is a familiar expedient in
statistical mechanics, for instance in the study of optimization
problems with simulated annealing and tempering.
One obtains here an interesting picture for all real
values of $\beta$ that fixes, with an uncertainty
that becomes smaller for larger systems, the value of the purity
of the subsystem under consideration, thus identifying an
``isoentangled" submanifold of states.
A strategy similar
to the one adopted here
was used in \cite{onebipartition} for the simpler case of bipartite
entanglement at a fixed bipartition, where the purity exhibits a
phase transition. We have seen that the
multipartite version of the problem is much more complicated, but it
would be of great interest
to understand whether the phase
transition that occurs in the bipartite situation, when there is no average over the
bipartitions, has a counterpart in the multipartite scenario.  This would
help elucidate the relation between multipartite entanglement,
complexity and the presence of frustration.

\ack This work is partly supported by the European Community
through the Integrated Project EuroSQIP.


\end{document}